# Capacitively-Induced Free-Carrier Effects in Nanoscale Silicon Waveguides for Electro-Optic Modulation

RAJAT SHARMA,[1,2] MATTHEW W. PUCKETT,[1,2]* HUNG-HSI LIN[2], ANDREI ISICHENKO[3], FELIPE VALLINI[2], AND YESHAIAHU FAINMAN[2]

[1]Equal contributors
[2]Department of Electrical & Computer Engineering, University of California, San Diego, 9500 Gilman Dr, La Jolla, CA 92023
[3]School of Applied and Engineering Physics, Cornell University, Ithaca, NY, 14853
*Corresponding author: mwpucket@ucsd.edu

We fabricate silicon waveguides in silicon-on-insulator (SOI) wafers clad with either silicon dioxide, silicon nitride, or aluminum oxide, and by measuring the electro-optic behavior of ring resonators, we characterize the cladding-dependent and capacitively-induced free-carrier effects in each of these waveguides. By comparing our measured data with simulation results, we confirm that the observed voltage dependencies of the transmission spectra are due to changes in the concentrations of holes and electrons within the semiconductor waveguide, and we show for the first time how strongly these effects depend on the cladding material which comes into contact with the silicon waveguide. Additionally, the waveguide loss is found to have a particularly high sensitivity to the applied voltage, and may therefore find use in a wide range of applications which require low- or high-loss propagation. Collectively, these phenomena may be incorporated into more complex waveguide designs in the future to create high-efficiency electro-optic modulators.

In recent years, silicon has become the primary candidate for the advancement of integrated photonics due to its prevalence within the electronics industry. One of the material's most notable shortcomings, however, is its centrosymmetry, which causes it to lack a second-order nonlinear susceptibility and consequently disallows electro-optic modulation based on the Pockels effect [1]. To circumvent this complication, a vast majority of research efforts involving electro-optic modulation in silicon waveguides have instead exploited the free-carrier plasma dispersion effect, in which a change in the concentration of free holes and electrons, generated by an electrical current, leads to deviations in both the real and imaginary parts of a semiconductor's index of refraction [2-4]. Additionally, some work over the past decade has been devoted to exploring the so-called strain-induced second-order nonlinear susceptibility in silicon [5-10]. By deforming silicon's diamond lattice in an asymmetric way, it is possible to remove the material's centrosymmetry, thereby generating a nonzero second-order nonlinearity within the material [5]. In recent work, values as high as 330 pm/V have been reported for the $\chi^{(2)}$ coefficient in strained silicon waveguides [6], making it a strong candidate in the design of nonlinear optical devices.

Recently, however, it has been found that strained silicon's electro-optic effect is roughly quadratic in nature, rather than linear, as would be expected for the Pockels effect [7]. Furthermore, many of the demonstrations of strained silicon's nonlinear properties have incorrectly assumed that the driving electric field used to control silicon's index of refraction penetrates strongly into the semiconductor waveguide itself [6-10], and this is now known to have led to inaccurate reported values of the nonlinear coefficient, $\chi^{(2)}$. Instead, the observed electro-refractive behavior in strained silicon waveguides is currently thought to be due to the capacitively-induced free-carrier effect [11-16].

In the following, we verify previously reported theoretical results which highlight the sensitivity of this effect to the material used to clad a silicon waveguide. By measuring the voltage dependence of the optical properties of silicon ring resonators coupled to bus waveguides, we show that the densities of fixed charges and interface traps at semiconductor-dielectric interfaces play a dominant role in determining how strongly the waveguides respond to bias voltages in terms of the effective indices of their supported modes. Based on the observed electro-optic behavior, we assert that capacitively-induced free-carrier effects in waveguides with appropriately chosen cladding layers may be used to realize high-efficiency optical modulators.

In our past work, we modeled the effects of free carriers in silicon waveguides clad with either silicon dioxide, silicon nitride, or aluminum oxide using the semiconductor physics tool Silvaco in combination with the finite-difference time-domain solver Lumerical [17,18]. We found that the real and imaginary parts of the TE-like mode's effective index changed differently with the applied bias voltage for each case, and were most sensitive to the voltage for the case of aluminum oxide. This was determined to be due to the material's high negative fixed charge density at the silicon interface, which drove the semiconductor waveguide into accumulation. To confirm that this behavior exists in reality, we fabricated silicon waveguides clad with silicon dioxide, silicon nitride, and aluminum oxide, then characterized the waveguides' electro-optic behavior by applying vertical bias voltages across ring resonators coupled to bus waveguides and characterizing their transmission spectra.

To fabricate our waveguides, we began with three silicon-on-insulator (SOI) wafers consisting of a 250 nm-thick device layer and a 3 μm-thick buried oxide layer. We spin coated the Dow Corning electron-beam resist XR-1541-006, which consists of dilute hydrogen silsesquioxane (HSQ), onto our wafers at a spin rate of 3000 rpm for 1 minute, and after spinning we baked our samples at 190ºC for 2 minutes. We then exposed the wafers to patterns corresponding to our waveguides using a Vistec EBPG 5200 electron-beam lithography system with a dose of 3500 μC/cm$^2$, and subsequently removed the unexposed resist through development for 1 minute in a solution of 1:4 tetramethylammonium hydroxide (TMAH). The uncovered silicon device layer was then removed in each of the wafers using an Oxford Plasmalab 100 Reactive Ion Etcher, and the gas flow rates used during the etch were 25 sccm for $SF_6$ and 50 sccm for $C_4F_8$. The chamber pressure was maintained throughout the etch at approximately 19 mTorr. After etching, the wafers were submerged for 10 seconds in a 1:10 buffered oxide etchant (BOE) solution and then rinsed thoroughly with deionized water to remove the remaining HSQ from the waveguides [19]. This step was critical because baked HSQ acts as a low-quality dielectric, the presence of which can reduce the reproducibility and reliability of electro-optic measurements [20]. Overhead view of the resulting device layout are shown in Fig. 1. Specifically, the images show a ring resonator coupled to a bus waveguide.

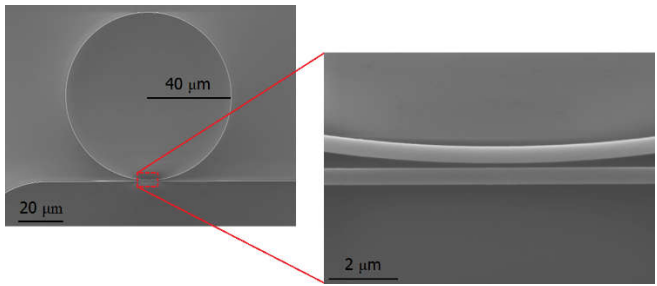

Fig. 1. SEM micrographs showing an unclad silicon ring resonator coupled to a bus waveguide. The ring radius is 40 μm, the waveguide width is 500 nm, and the separation between the bus waveguide and the ring is 100 nm.

One of the three samples was then clad with 50 nm of silicon nitride using an Oxford Plasmalab 100 Plasma-Enhanced Chemical Vapor Depositor (PECVD), whereas a second was clad with 50 nm of aluminum oxide using a Beneq TFS200 Atomic Layer Depositor (ALD). The resulting waveguide cross-section is shown for the silicon nitride case in Fig. 2a. Such thin layers of silicon nitride and aluminum oxide were chosen because our intent was to change the electrical properties of the semiconductor-dielectric interfaces without dramatically altering the waveguides' optical properties. To improve the film quality and reduce the density of interface traps, a rapid thermal anneal (RTA) was carried out for both the silicon nitride and aluminum oxide wafers for 15 minutes at a temperature of 300ºC in a forming gas ambient ($N_2/H_2$:90%:10%) [21]. A 1.5 μm-thick silicon dioxide cladding layer was then deposited on all three samples, again using PECVD, and the rapid thermal anneal was repeated. The resulting TE-like optical modes supported by the three waveguide geometries are shown in Fig. 2b through 2d. It is important to note that the magnitude of the fixed charge present at the semiconductor-dielectric interfaces, and hence the magnitude and sign of the electro-optic effect, may be controlled in the future, for each of the given cladding layers, by modifying either the deposition conditions, the annealing conditions, or the wafer pretreatments [21,22]. After the fabrication of the waveguides was complete, photolithography was carried out to create aluminum electrodes above the ring resonators, and the samples were finally diced to expose the waveguides' end facets and allow butt-coupling.

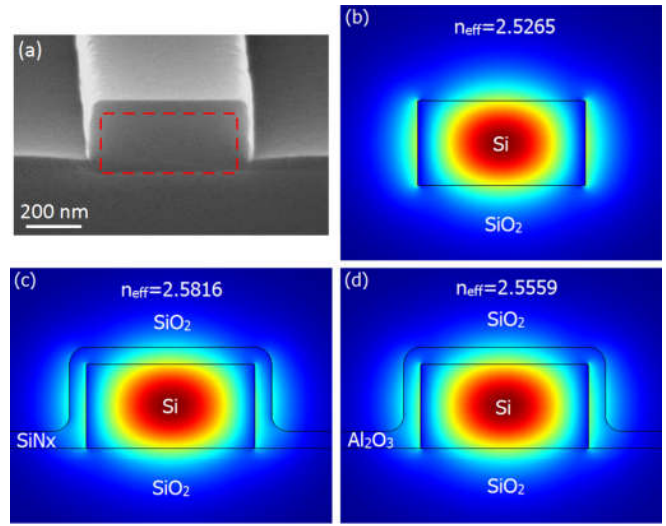

Fig. 2. (a) SEM micrograph of a 500 nm-wide, 250 nm-tall silicon waveguide (red dashed line) clad with 50 nm of silicon nitride. (b-d) The supported TE-like mode at λ=1.55 μm for the same silicon waveguide clad with (b) 50 nm of silicon nitride followed by silicon dioxide, (c) 50 nm of aluminum oxide followed by silicon dioxide, and (d) silicon dioxide only. The total variation in effective index is less than 0.06, which is approximately 2.4% of the value for the aluminum oxide case.

As in past work, we used Silvaco to model the electrical properties of our three waveguide types, using previously reported values for the fixed charge and interface trap densities characteristic of each cladding material [12]. In these models, the doping of the silicon device layer was set to p-type, assuming a boron dopant concentration of $10^{15}$ cm$^{-3}$. We generated spatially resolved, voltage-dependent maps of the electron and hole concentrations within the waveguides, then translated these values through the Soref and Bennet equations into local changes in silicon's index of refraction [23]. It should be noted that recent work has found the Soref and Bennet equations tend to underestimate the electro-absorption effect due to gate induced charge layers [24]. However, this discrepancy only becomes significant for values of the local electric field exceeding $10^8$ V/m, and such high fields were not observed for the limited range of applied voltages used in our models. Additionally, silicon nitride films deposited on silicon have been shown experimentally to possess values of fixed charge which change in response to applied bias voltages, but this non-ideality is only significant for voltages beyond those considered here [25].

By combining our results with the software Lumerical, we were able to predict how the effective indices of the waveguides' TE-like modes would change with voltage. Our results, shown in Fig. 3, are in agreement with the results of previous work, showing that the slopes of both the real and imaginary index curves change sign for the case of aluminum oxide, in comparison to the silicon nitride and silicon dioxide cases. As we have previously shown, the electric field changes only marginally within the waveguide in response to bias voltages, and this eliminates the possibility of any appreciable strain-induced Pockels effect occurring [12]. The free-carrier effect simulated here is then anticipated to dominate the overall electro-optic behavior.

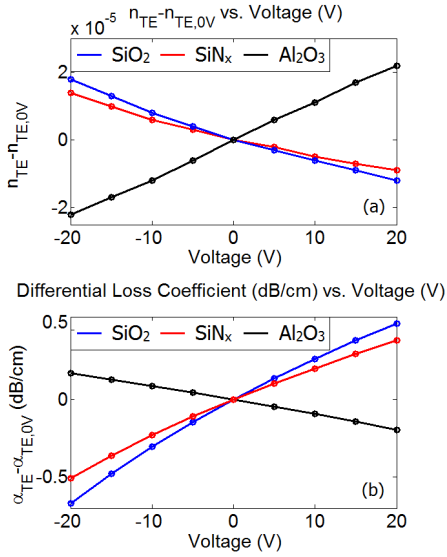

Fig. 3. Theoretical electro-optic characteristics in terms of (a) the effective index ($n_{TE}$) and (b) the loss coefficient ($\alpha_{TE}$) for silicon waveguides clad with either silicon dioxide (blue), silicon nitride (red), or aluminum oxide (black).

To experimentally measure these effects in our waveguides, we employed the optical setup shown schematically in Fig. 4. An infrared laser, tunable from 1465 to 1575 nm, was sent via an optical fiber into a polarization scrambler and a tunable polarizer, then coupled into our samples at a horizontal (TE) polarization using a lensed tapered fiber. The light emitted from the samples was collected with a metallic output objective, and the mode was magnified using two 4F systems of lenses. The power transmitted through the waveguides was measured using an optical power meter, and the transmission spectrum was obtained by incrementally increasing the emission wavelength of the laser. To apply a bias voltage across our samples, we touched probe tips to both the aluminum electrodes and the aluminum plate on which the samples were mounted, then connected the probes to a power source ranging from negative to positive 20 V.

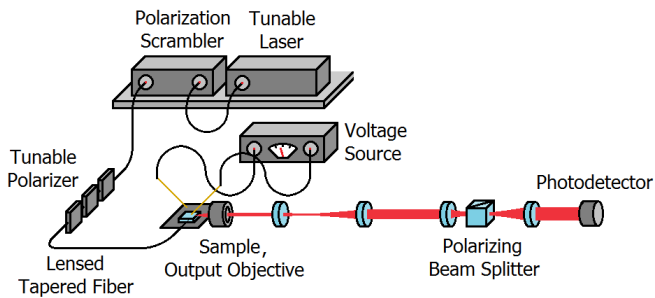

Fig. 4. Illustration of the experimental setup used to characterize the electro-optic properties of our fabricated waveguides.

The transmission spectra we measured exhibited periodic dips at each of the rings' resonant wavelengths, and the shapes and positions of these dips could be fit to Lorentzian curves in order to extract the waveguides' loss coefficients and refractive indices. Specifically, the resonant transmission dip could be fit to the expression [26]:

$$T = \frac{t^2 - 2t\tau \cos(\theta) + \tau^2}{1 - 2t\tau \cos(\theta) + (t\tau)^2} \quad (1)$$

where t is the self-coupling coefficient, $\tau$ is the attenuation coefficient, and $\theta$ is the single-pass phase shift, in turn defined as [26]:

$$\theta = \frac{4\pi^2}{\lambda} n_{eff} r \quad (2)$$

where $n_{eff}$ is the effective index of the mode, r is the ring radius, and $\lambda$ is the unbiased resonant wavelength. Once the value of $\tau$ was obtained from the numerical fit, it could be used to derive the loss coefficient as [26]:

$$\alpha = \frac{20 \log_{10}(\tau)}{2\pi r} \quad (3)$$

Additionally, the change in the real part of a mode's effective index induced by the bias voltage could be calculated from the spectral shift of the optical resonance as [7]:

$$\Delta n_{eff} = \frac{\Delta \lambda}{\lambda} n_{eff} \quad (4)$$

where $\Delta\lambda$ is the observed change in the central wavelength of the resonance and $n_{eff}$ is the unbiased effective index.

Fig. 5a through 5c show single resonances for the silicon dioxide, silicon nitride, and aluminum oxide clad samples, and the passive loss coefficients extracted from these data were 11.3, 9.3, and 9.9 dB/cm, respectively, while Fig. 5d shows the voltage dependent transmission spectra for the case of $SiO_2$. Additionally, Fig. 6a and 6b show the results of the electro-optic measurements, plotting the changes in both the real part of the effective index and the loss coefficient. These results are in good agreement with the theoretically produced values, although the magnitude of the electro-refractive effect for the case of the silicon nitride cladding is found to be significantly larger than predicted. This is most likely due to a discrepancy in the values of the fixed charge between the two cases, resulting from the unique processing steps the silicon waveguide's surfaces were subjected to prior to the PECVD of the cladding layer. Also, the loss is seen to be much more sensitive to the applied voltage than theoretically predicted for each of the three geometries. This may be due to uncertainties in (1) the interface trap density at the silicon-dielectric interfaces, (2) the index of refraction of the cladding layers, or (3) dissimilar values of fixed charge between the upper wall and the sidewalls of the silicon waveguide, due to the different set of processes and cleaning steps that they have been subjected to, which have not been considered in our simulations as a possibility.

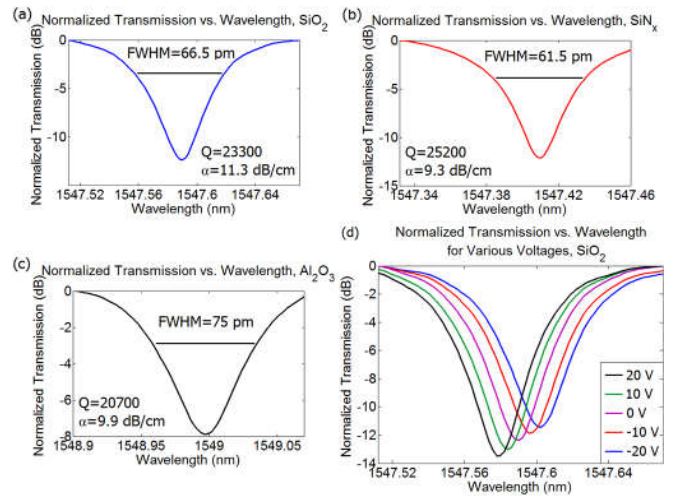

Fig. 5. Passive transmission spectra showing single resonances for rings in (a) silicon dioxide, (b) silicon nitride, and (c) aluminum oxide clad waveguides, (d) Voltage-dependent transmission spectrum for the case of $SiO_2$

It is also important to note that none of the cladding layers used in this work contained appreciable stress, further invalidating the strain-induced nonlinearity as a possible source of the observed electro-optic effects. The stresses contained within the aluminum oxide, silicon nitride, and silicon dioxide cladding layers were measured using a Toho Technology FLX-2320 Thin Film Stress Measurement System to be approximately 200 MPa, -300 MPa, and 200 MPa, respectively, and these values are nearly an order of magnitude smaller than those used to fabricate strained silicon waveguides [5-10].

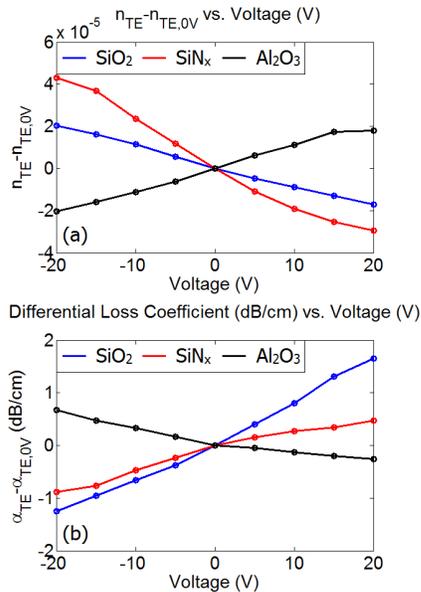

Fig. 6. Experimental electro-optic characteristics in terms of (a) the effective index ($n_{TE}$) and (b) the loss coefficient ($\alpha_{TE}$) for silicon waveguides clad with either silicon dioxide (blue), silicon nitride (red), or aluminum oxide (black).

In conclusion, we have experimentally shown how different dielectric claddings impact the capacitively-induced electro-optic effect in nanoscale silicon waveguides. Whereas the positive fixed charges present in the cases of silicon dioxide and silicon nitride drive p-type silicon into inversion, aluminum oxide's negative fixed charges have the opposite effect, bending the semiconductor's valence and conduction bands in the opposite direction and inducing accumulation. Consequently, the sign of the electro-optic effect is reversed for the case of the latter. We believe this result is important because it clearly indicates how the capacitively-induced free-carrier effect may, in waveguides with properly chosen cladding materials and consequent fixed charges, be leveraged to realize CMOS-compatible electro-optic modulators which do not rely on the injection of electrical currents to control a waveguide's optical characteristics. Additionally, we would like to highlight the fact that a dielectric-clad silicon waveguide's optical losses may be reduced or increased by applying an appropriately chosen bias voltage, allowing for the in-vivo tuning and optimization of a wide variety of photonic devices.

**Funding.** The National Science Foundation (NSF); the Defense Advanced Research Projects Agency (DARPA); the NSF ERC CIAN; the Office of Naval Research (ONR); the Multidisciplinary University Research Initiative (MURI); the Cymer corporation.

**Acknowledgment**. We thank UCSD's Nano3 cleanroom staff, namely Dr. Maribel Montero and Mr. Ryan Anderson, for their assistance with sample fabrication and imaging.